\documentclass[a4paper]{llncs}

\usepackage{graphicx}
\usepackage[font={small}]{caption}
\usepackage{subcaption}
\usepackage{amsmath, amssymb}
\usepackage{hyperref}
\usepackage[colorinlistoftodos]{todonotes}

\usepackage{tikz}
\tikzstyle{vertex}=[circle,minimum size=8pt,inner sep=0pt]
\tikzstyle{vertex2} = [draw,circle,minimum size=3pt,inner sep=0pt]
\tikzstyle{vertex3} = [draw,circle,minimum size=8pt,inner sep=0pt]

\newtheorem{thm}{Theorem}
\newtheorem{define}[thm]{Definition}

\title{Control Explicit---Data Symbolic\\Model Checking: An
  Introduction\thanks{This work was supported by the Czech Grant Agency
    grant No. GAP202/11/0312.}}

\author{Ji{\v r}{\'i} Barnat and Petr Bauch}

\institute{Faculty of Informatics, Masaryk University\\
  Brno, Czech Republic\\
  \{barnat, xbauch\}@fi.muni.cz}

\begin{document}

\maketitle

\begin{abstract}
  A comprehensive verification of parallel software imposes three crucial
  requirements on the procedure that implements it. Apart from accepting real
  code as program input and temporal formulae as specification input, the
  verification should be exhaustive, with respect to both control and data
  flows. This paper is concerned with the third requirement, proposing to
  combine explicit model checking to handle the control with symbolic set
  representations to handle the data. The combination of explicit and symbolic
  approaches is first investigated theoretically and we report the requirements
  on the symbolic representation and the changes to the model checking process
  the combination entails. The feasibility and efficiency of the combination is
  demonstrated on a case study using the DVE modelling language and we report a
  marked improvement in scalability compared to previous solutions. The results
  described in this paper show the potential to meet all three requirements for
  automatic verification in a single procedure combining explicit model checking
  with symbolic set representations.
\end{abstract}

\section{Introduction} \label{sec:int}

Specification of the intended behaviour of a computer system forms the basis of
any rigorous, contract-based development. The final product must comply with its
specification, and until it does, until all functional, safety and performance
requirements are met, the development continues and the expenses
grow. Requirements on safety and performance are rarely formalised, and thus
compliance with those requirements is commonly ensured by strictly adhering to an
established set of rules, e.g. DO-178B~\cite{RCT92} for aviation
systems. Functional requirements, on the other hand, often can be expressed in a
precise, formal language -- a property that makes them amenable to verification
using formal methods.

Not all formal methods currently in practice, however, can handle requirements
formalised in a language of sufficient expressivity. When programs behave
nondeterministically, when they react to unpredictable environment or when the
interleaving of components executed in parallel is unknown, the developers often
need to express the desired behaviour as it evolves in time, using temporal
logics~\cite{MP81}. Among the \emph{execution-based verification} methods that
exist, i.e. \emph{testing}, \emph{symbolic execution}~\cite{Kin76} and
\emph{model checking}~\cite{CES86}, only model checking is able to verify
that a system is a model of the required temporal property~\cite{BDK08}.

Yet model checking in the present state is far from replacing testing and
symbolic execution in real-world application. Apart from the well-known and
well-understood problem of state space explosion, there are other aspects that
prevent more widespread use. The one addressed by this paper is the restriction
to \emph{closed systems}, i.e. programs where each variable is initialised to a
fixed value (see the related work for more detailed discussion of other attempts
at model checking of open systems). Symbolic execution is not limited to closed
systems; there the values of variables are represented symbolically, which in
theory enables all possible values to be considered within a single run of such
execution.

It seems that unifying these two formal methods, symbolic execution to gain
access to open systems and model checking for temporal properties, could lead to
a method of high practical value. Indeed, for carrying out unit
tests~\cite{Run06} on a nondeterministic component within a larger program,
neither technique alone would suffice to achieve substantial reliability of the
product. Our approach to this unification is to augment the model checking by
allowing to verify the correctness for multiple variable evaluations at a
time. To represent the state space explicitly, but with the states being in fact
symbolic sets of states: hence \emph{control explicit---data symbolic model
  checking}.

A straightforward way for supporting open variables is to repeat the
verification, as many times as there are combinations of the input values. It
requires only a very small change in the implementation, as was demonstrated
in~\cite{BBB12c}, where the model checking was suitably modified for
verification of Simulink circuits. When generating successors of a circuit
state, the process first takes into consideration the branching in the
specification transition system and then the branching in the program transition
system (the control-flow branching). Resolution of the data-flow branching
(caused by open input variables) can be attached to that of control-flow or,
since Simulink circuits are otherwise deterministic, replace it. In other words,
the states of the circuit model are treated as if having one successor for every
combination of the input variable evaluations.

Needless to say, this approach is extremely demanding with respect to
computational resources. Apart from the original (often merely potential)
explosion caused by control-flow nondeterminism there is now an addition,
inescapable explosion of data. Every variable multiplies the number of
successors of every state by the size of its range. Even if the circuit only
used Boolean variables, or equivalently if the range was always equal to two,
the blow-up would still be exponential. For explicit model checking especially,
such an approach is an interesting proof of concept, baseline for future
improvements, but limited in application to academic examples and
experiments. For use in practice, e.g. in industry-level unit testing, a
cleverer approach needs to be adopted.

\paragraph{Contribution}

On the most fundamental level, the modification of explicit model checking
proposed in this paper lies in replacing the exponential number of states in the
transition system with more complex successor generation. Model checking systems
with input variables, i.e. with nontrivial data flow, entails either further
state space explosion or employing some form of symbolic
representation. Throughout this paper, we propose representing symbolically only
the data part; the control part remains explicit. Not every symbolic
representation, however, can be used and we detail as to what requirements must
the representation meet to enable model checking against temporal
properties. Using a basic representation that meets the proposed requirements,
we have described how the model checking process must be modified to represent
data symbolically. The experiments on Peterson's communication protocol report
far better scalability compared to the purely explicit approach. Replacing one
level of the state space explosion for complex symbolic states and for the
additional difficulty associated with their generation appears to have the
potential to forward the progress towards practical verification of concurrent
systems.

\subsection{Related Work}

Of the plethora of papers pertaining to execution-based verification only a few
are directly related to the presented work. Firstly, there is the symbolic
execution and the related research aiming at improving its robustness. For
example, support for parallel or otherwise nondeterministically behaving systems
was first incorporated in~\cite{KPV03}. Allowing specification in LTL was
partially introduced in~\cite{BDK08}, yet the undecidability of state matching
limited the approach to only a small subset of LTL. The research that perhaps
most closely resembles ours was described in a section on \emph{Delayed
  Nondeterminism} in a PhD Thesis by Schlich~\cite{Sch08}. There the variables
were represented symbolically until used and then the algorithm opted for the
explicit representation.

Symbolic model checking is most commonly applied on Boolean programs, avoiding
many of the mentioned problems, especially those related to
arithmetic. Computing multiplication with the standard representation, Binary
Decision Diagrams~\cite{McM92}, is exponential in the size of the
representation~\cite{Bry91}. Other representation were designed to remedy this
deficiency, such as Binary Moment Diagrams~\cite{BC95} or Boolean Expression
Diagrams~\cite{WBC00}. These represent variables on the word level rather then
on the binary level. Another direction of research attempted to utilise the
advance of modern satisfiability solvers, first with classical SAT~\cite{BCC99}
and then with the more specific SMT~\cite{AMP06}. However, SAT-based model
checkers allow the state space to be traversed only to a bounded depth, which
renders such model checking incomplete. It was also suggested to limit the
scope of the symbolic model checking to programs with only Presburger
arithmetics, where more efficient representation were applicable, e.g. Periodic
Sets~\cite{BW94}.

Various combinations of different approaches and representations have been
devised and experimented with. When multiple representation were combined, it
was mostly to improve on weak aspects of either of the representations, for
example in~\cite{YWG06} multiple symbolic representation for Boolean and integer
variables were employed in combination.
Finally, the two approaches to model checking, explicit
and symbolic, were combined to improve solely upon control-flow
nondeterminism. Some improvement was achieved by storing multiple explicit
states in a single symbolic state~\cite{DLK11} or by storing explicitly the
property and symbolically the system description~\cite{STV05}.

Our stating that model checking is restricted to closed systems requires further
discussion. \emph{Module checking}, introduced in~\cite{KV96} and detailed
in~\cite{God03}, allows verification of open systems, though the meaning of
\emph{openness} differs from ours. The two sources of nondeterminism in module
checking are internal and external, where the external nondeterminism is
controlled by the environment. A system is open in the sense that the
environment may restrict the nondeterminism and the verification has to be
robust with respect to arbitrary restriction. The approach to verifying open
systems also differs since only branching time logics can distinguish open from
closed systems, in the module checking sense. For linear time logics, every path
has to satisfy the property and thus open and closed systems collapse into one
source of nondeterminism; where this paper intends to separate the
nondeterministic choices emerging from control and data flows.

Much closer to our separation between control and data is the work initiated by
Lin~\cite{Lin96}. Lin's \emph{Symbolic Transition Graphs with Assignments}
represent precisely parallel programs with input variables using a combination
of first-order logic and process algebra for communicating systems. Similarly as
for symbolic execution, the most complicated aspect of this representation is
the handling of loops. Lin's solution computes the greatest fix point of a
predicate system representing the first-order term for each loop. Then two
transition graph are bisimilar if the predicate systems representing all loops
are equivalent. While the theoretical aspects of our work are very similar to
Lin's it is not clear how his equality of predicate systems could be used in LTL
model checking, though it is intended as one of our future directions in
research.

Finally, our work can be seen as an alternative approach to that described
in~\cite{HGD95}. There the authors also divide parallel programs into control
and data, where control is handled using symbolic model checking and data by
purely symbolic manipulation of first-order formulae. Avoiding the problems with
loop -- which were the main objective of Lin's work -- by not allowing symbolic
data to influence control, the authors of~\cite{HGD95} implemented verification
of parallel programs against first-order branching logic. Hence while their
distinction between control and data is almost precisely equivalent to ours, the
method proposed in this paper allows verification against linear time logic with
no restriction on the parallel program. The loops still pose a considerable
problem, but can only severely increase the running time; they never render the
verification task undecidable.

\section{Preliminaries} \label{sec:pre}

The methodology proposed in this paper depends on various technical aspects of
explicit model checking and specific input languages. These must be at least
generally described for the purpose of further discussion. Within this section
we will proceed from the more theoretical to more practical, from the
foundations of model checking to its implementation.

\begin{define} \label{def:ltl}
  Let $\Sigma$ be the set of \emph{atomic propositions}. Then this recursive
  definition specifies all well-formed \emph{LTL formulae} over $\Sigma$, where
  $p\in \Sigma$:
  \begin{center}
    $\Psi::=p\mid\neg\Psi\mid\Psi\wedge\Psi\mid X\ \Psi\mid\Psi\ U\ \Psi$
  \end{center}
\end{define}

\begin{example} \label{exa:ltl}
  There are some well-established syntactic simplifications of the LTL language,
  e.g. $\mathit{false}:=p\wedge\neg p$, $\mathit{true}:=\neg \mathit{false}$,
  $\phi\Rightarrow\psi:=\neg(\phi\wedge \neg\psi)$, $F\ \phi:=$
  $\mathit{true}\ U\ \phi$, $G\ \phi:=\neg (F\ \neg\phi)$. Assuming that
  $\Sigma=\{\alpha:=(c=5),\beta:=(a\neq b)\}$, these are examples of well-formed
  LTL formulae: $G\ \beta, \alpha\ U\ \neg\beta$. Informally, the first one
  states that $a$ must never be equal to $b$ and the second that $c$ is equal to
  $5$ as long as $a$ equals $b$ (and at some point $a$ must become different
  from $b$).\hfill$\triangle$
\end{example}

\begin{define} \label{def:lts}
  A \emph{Labelled Transition System (LTS)} $M=(S, \rightarrow, \nu, s_0)$ is a
  tuple, where: $S$ is a set of states, $\rightarrow\subseteq S\times S$ is a
  transition relation, $\nu:S\rightarrow 2^{\Sigma}$ is a valuation function and
  $s_0\in S$ is the initial state. A function $r:\mathbb{N}\rightarrow S$ is an
  \emph{infinite run} over the states of $M$ if $r(0)=s_0, \forall i:
  r(i)\rightarrow r(i+1)$. The \emph{trace} or \emph{word} of a run is a
  function $w:\mathbb{N}\rightarrow 2^{\Sigma}$, where $w(i)=\nu(r(i))$.
\end{define}


Traversing an LTS requires the underlying graph to be represented, in some form,
in the computer memory. There are two categories of suitable graph
representations: explicit, where vertices and edges are already stored in the
memory and implicit, where successors are generated on-the-fly from the
description of their predecessors. For implicit representation, only two
functions must be provided as the system description: \textsf{initial state} to
generate the initial system configuration and \textsf{successors}. The latter
function takes as the input a single state and, based on the control-flow
choices available in that state, returns the set of successor states of the
input state.

An LTL formula states a property pertaining to an infinite trace; see how traces
relate to runs in Definition~\ref{def:lts}. Assuming the LTS is a model of a
computer program then a trace represents one specific execution of the
program. Also the infiniteness of the executions is not necessarily an error --
programs such as operating systems or controlling protocols are not supposed to
terminate (and indeed would be incorrect if they did terminate).

\begin{define} \label{def:sem}
  Let $w$ be an infinite word and let $\Psi$ be an LTL formula over
  $\Sigma$. Then the following rules decide if $w$ satisfies $\Psi$,
  $w\models\Psi$, where $w(i)$ is the $i$-th letter of $w$ and $w_i$ is the
  $i$-th suffix of $w$:
  \[
  \begin{array}{lcl}
    w\models p &\ \mathit{iff} &\ p\in w(0),\\

    w\models \neg\Psi &\ \mathit{iff} &\ w\not\models\Psi,\\

    w\models \Psi_1\wedge\Psi_2 &\ \mathit{iff} &\ w\models \Psi_1
    \ \mathit{and}\ w\models \Psi_2,\\

    w\models X\ \Psi &\ \mathit{iff} &\ w_1\models\Psi,\\

    w\models \Psi_1\ U\ \Psi_2 &\ \mathit{iff} &\ \exists i\forall j<i: w_j\models
    \Psi_1, w_i\models\Psi_2.
  \end{array}
  \]
\end{define}

Clearly, a system as a whole satisfies an LTL formula if all its executions (all
infinite words over the states of its LTS) do. Efficient verification of that
satisfaction, however, requires a more systematic approach than enumeration of
all executions. An example of a successful approach is the enumerative approach
using \emph{B{\"u}chi automata}.

\begin{define} \label{def:ba}
  A \emph{B{\"u}chi automaton} is a tuple $A=(S, \rightarrow, \nu, s_0, F)$,
  formed of an LTS $M=(S,\rightarrow,\nu,s_0)$ and $F\subseteq S$. An automaton
  $A$ accepts an infinite word $w$, $w\in L(A)$, if there exists a run $r$ for
  $w$ in $M$ and there is a state from $F$ that appears infinitely often on $r$,
  i.e. $\forall i\exists j>i: r(j)\in F$.
\end{define}

Arbitrary LTL formula $\phi$ can be transformed into a B{\"u}chi automaton
$A_{\phi}$ such that $w\models\phi\Leftrightarrow w\in L(A_{\phi})$. Also
checking that every execution satisfies $\phi$ is equivalent to checking that no
execution satisfies $\neg\phi$. It only remains to combine the LTS model of the
given system $M$ with $A_{\neg\phi}$ in such a way that the resulting automaton
will accept exactly those words of $M$ that violate $\phi$. Finally, deciding
existence of such a word -- and by extension verifying correctness of the system
-- is equivalent to detecting accepting cycle in the underlying
graph~\cite{CES86}.

Returning back to the implementation of the model checking process, already
having the \textsf{initial state} and \textsf{successors} functions, the
description can be finalised by adding one more function: \textsf{is
  accepting}. With this function, which returns a binary answer when provided
with an LTS state, one can represent B{\"u}chi automata and consequently detect
accepting cycles within.

That is, however, a significant step and not entirely a trivial one. A state of
the product LTS comprises two states, one for the specification and one for the
system. It follows that the \textsf{successors} function must also be modified,
because now there are control choices also from the specification LTS. These
control choices are based on whether or not particular atomic propositions hold
in the input state, such that the property remains satisfied. The described
modifications are sufficient for LTL model checking, assuming, and that is an
aspect of major importance for this paper, that the states of the product LTS
are stored and that duplicates in the state space can be detected. The basis for
this importance will become apparent in the next chapter.

\section{Explicit Control with Symbolic Data}

Automata-based model checking, as presented in the previous section, handles
only control-flow nondeterminism. That would be perfectly sufficient if
communication protocols were the only type of input models, but should model
checking aspire to verify correctness of real software, such limitation would
decrease its usability. Small units of programs often take inputs and possible
values of these inputs -- each defining a new, potentially unique execution --
must be considered as well; otherwise the verification would not be exhaustive.

Handling of both sources of nondeterminism combined within a single procedure is
a logical next step when adapting model checking for the use in unit
testing. This paper proposes allowing the specification of ranges of the input
variables in verified programs, i.e. allowing verification with open variables,
even though bounded. Two approaches for handling such modification present
themselves. Firstly, simply run one instance of the model checking process for
every combination of the input variables: an approach described in
Section~\ref{sec:int}. Secondly, and what is devised in this paper, run model
checking only once but instead of simple, single-value states use
\emph{multi-states} encoding multiple values of variables.

\subsection{Set-Based Reduction}

The states of computation in a parallel program are uniquely defined by the
evaluation of variables and the program counters of individual threads. Other
program components needed for execution, such as stack and heap contents, are
assumed to be represented as fresh variables. Given this abstraction we can
define a transition system generated by the execution of a parallel program in
exactly the same manner as in Definition~\ref{def:lts}. For the purposes of
distinguishing the two sources of nondeterminism, control and data, we will
associate with a parallel program $P$, a transition system $\mathcal{T}^P=
(S,\rightarrow,\nu,S_0)$, where each $s\in S$ is composed of two parts
$s=(s_c,s_d)$. (Also $S_0$ is a subset of $S$, since there generally are many
initial evaluations.)  There $s_c$ represents the evaluation of program counters
and other variables that are not modified externally and $s_d$ represents the
evaluation of input variables. Similar state composition is preserved when the
product with a B{\"u}chi automaton is computed, i.e. given a program $P$ and a
B{\"u}chi automaton $A_{\varphi}$, the states of the product
$\mathcal{A}^P_{\varphi}$ are again composed of two parts, where the information
identifying the states of $A_{\varphi}$ is part of $s_c$.

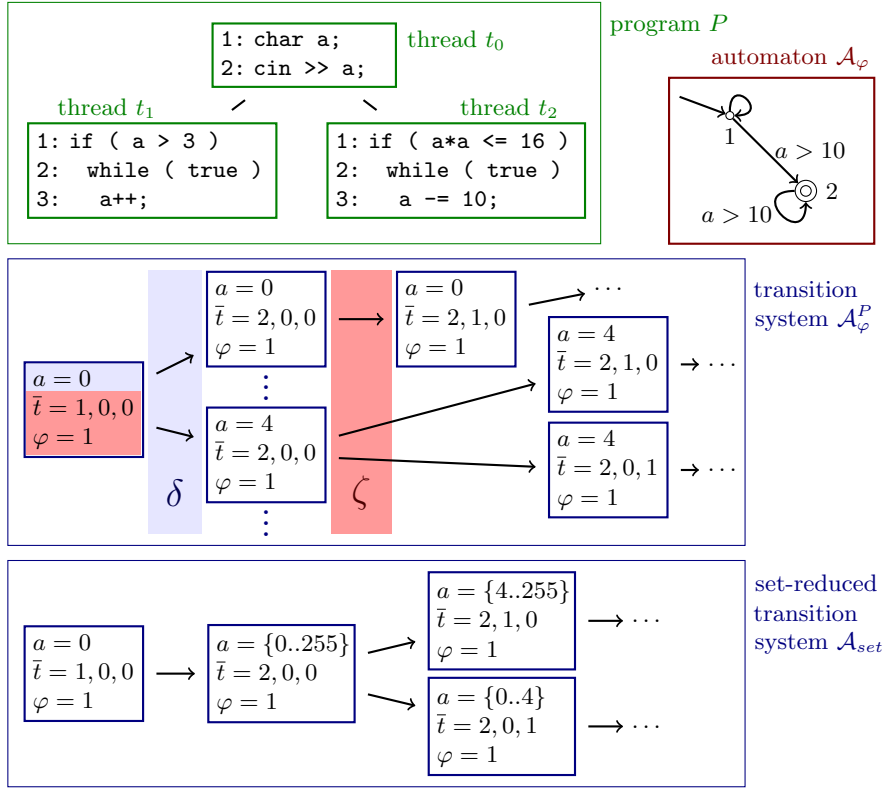
\begin{figure}[t]
  \centering
  \begin{tikzpicture}
    \node (A) at (4,2) {\textcolor{green!50!black}{thread $t_0$}};
    \node (A) at (-0.6,1.1) {\textcolor{green!50!black}{thread $t_1$}};
    \node[draw=green!50!black,thick,rectangle,outer sep=5pt]
    (th0) at (2,1.8) {
      \begin{minipage}{0.18\textwidth}
        \texttt{\footnotesize{
            \hspace*{-0.5em}1:\hspace*{-0.2em} char a;\\
            2:\hspace*{-0.2em} cin >> a;
        }}
      \end{minipage}
    };
    \node[draw=green!50!black,thick,rectangle,outer sep=5pt]
    (th1) at (0,0.27) {
      \begin{minipage}{0.25\textwidth}
        \texttt{\footnotesize{
            \hspace*{-0.5em}1:\hspace*{-0.2em} if ( a > 3 )\\
            2:\hspace*{0.5em} while ( true )\\
            3:\hspace*{1em} a++;
        }}
      \end{minipage}
    };
    \node (A) at (4.7,1.1) {\textcolor{green!50!black}{thread $t_2$}};
    \node (A) at (6.8,2.2) {\textcolor{green!50!black}{program $P$}};
    \node[draw=green!50!black,thick,rectangle,outer sep=5pt]
    (th2) at (4,0.27) {
      \begin{minipage}{0.26\textwidth}
        \texttt{\footnotesize{
            \hspace*{-0.5em}1:\hspace*{-0.2em} if ( a*a <= 16 )\\
            2:\hspace*{0.5em} while ( true )\\
            3:\hspace*{1em} a -= 10;
        }}
      \end{minipage}
    };
    \path[draw=green!50!black] (-1.9,-0.7) rectangle (5.9,2.5);
    \path[draw,thick] (th0) -- (th1);
    \path[draw,thick] (th0) -- (th2);
    \node (A) at (8.4,1.75) {\textcolor{red!50!black}{automaton
        $\mathcal{A}_{\varphi}$}};
    \path[draw=red!50!black,thick] (6.8,-0.7) rectangle (9.5,1.5);
    \node[vertex] (e) at (6.8,1.3)  {};
    \node[vertex2] (f) at (7.6,1) [label=below:{\footnotesize $1$}] {};
    \node[vertex3] (g) at (8.6,0) [label=right:{\footnotesize $2$}] {};
    \draw (8.6,0) circle(2pt);
    \path[draw,thick,->] (e) -- (f);
    \path[draw,thick,->] (f) -- (g) node [midway,right] {$a>10$};
    \path[draw,thick,->] (f) to [in=0,out=90,min distance=10mm,loop] (f);
    \path[draw,thick,->] (g) to [in=270,out=180,min distance=10mm,loop] node
         [midway,left] {$a>10$} (g);
    \path[fill=red!40] (-1.65,-3.5) rectangle (-0.15,-2.3);
    \path[fill=blue!10] (-1.65,-2.65) rectangle (-0.15,-2.3);
     \node[draw=blue!50!black,thick,rectangle,outer sep=5pt]
     (top) at (-0.9,-2.9) {
       \begin{minipage}{0.11\textwidth}
         \footnotesize{
           $a=0$\\
           $\overline{t}=1,0,0$\\
           $\varphi=1$
         }
       \end{minipage}
     };
     \node[draw=blue!50!black,thick,rectangle,outer sep=5pt]
     (l1) at (1.5,-1.7) {
       \begin{minipage}{0.11\textwidth}
         \footnotesize{
           $a=0$\\
           $\overline{t}=2,0,0$\\
           $\varphi=1$
         }
       \end{minipage}
     };
     \node[draw=blue!50!black,thick,rectangle,outer sep=5pt]
     (l2) at (1.5,-3.5) {
       \begin{minipage}{0.11\textwidth}
         \footnotesize{
           $a=4$\\
           $\overline{t}=2,0,0$\\
           $\varphi=1$
         }
       \end{minipage}
     };
     \node[draw=blue!50!black,thick,rectangle,outer sep=5pt]
     (l3) at (4,-1.7) {
       \begin{minipage}{0.11\textwidth}
         \footnotesize{
           $a=0$\\
           $\overline{t}=2,1,0$\\
           $\varphi=1$
         }
       \end{minipage}
     };
     \node[draw=blue!50!black,thick,rectangle,outer sep=5pt]
     (l4) at (6,-2.3) {
       \begin{minipage}{0.11\textwidth}
         \footnotesize{
           $a=4$\\
           $\overline{t}=2,1,0$\\
           $\varphi=1$
         }
       \end{minipage}
     };
     \node[draw=blue!50!black,thick,rectangle,outer sep=5pt]
     (l5) at (6,-3.7) {
       \begin{minipage}{0.11\textwidth}
         \footnotesize{
           $a=4$\\
           $\overline{t}=2,0,1$\\
           $\varphi=1$
         }
       \end{minipage}
     };
     \path[draw=blue!50!black] (-1.9,-4.7) rectangle (7.8,-0.9);
     \node (dots2) at (1.5,-2.5)
           {\textcolor{blue!50!black}{\textbf{$\vdots$}}};
     \node (dots2) at (1.5,-4.3)
           {\textcolor{blue!50!black}{\textbf{$\vdots$}}};
     \node (A) at (8.6,-1.3) {\textcolor{blue!50!black}{transition}};
     \node (A) at (8.71,-1.7) {\textcolor{blue!50!black}{system
         $\mathcal{A}^P_{\varphi}$}};
     \node (dots3) at (6,-1.3) {\textbf{$\ldots$}};
     \node (dots4) at (7.5,-2.3) {\textbf{$\ldots$}};
     \node (dots5) at (7.5,-3.7) {\textbf{$\ldots$}};
     \path[fill=blue!10] (-0.05,-4.55) rectangle (0.65,-1.05);
     \node (A) at (0.3,-4) {\Large\textcolor{blue!30!black}{$\delta$}};
     \path[fill=red!40] (2.35,-4.55) rectangle (3.15,-1.05);
     \node (A) at (2.75,-4) {\Large\textcolor{red!30!black}{$\zeta$}};
     \path[draw,thick,->] (top) -- (l1);
     \path[draw,thick,->] (top) -- (l2);
     \path[draw,thick,->] (l2) -- (l4);
     \path[draw,thick,->] (l2) -- (l5);
     \path[draw,thick,->] (l1) -- (l3);
     \path[draw,thick,->] (l3) -- (dots3);
     \path[draw,thick,->] (l4) -- (dots4);
     \path[draw,thick,->] (l5) -- (dots5);

     \node[draw=blue!50!black,thick,rectangle,outer sep=5pt]
     (top2) at (-0.9,-6.4) {
       \begin{minipage}{0.11\textwidth}
         \footnotesize{
           $a=0$\\
           $\overline{t}=1,0,0$\\
           $\varphi=1$
         }
       \end{minipage}
     };
     \node[draw=blue!50!black,thick,rectangle,outer sep=5pt]
     (l21) at (1.7,-6.4) {
       \begin{minipage}{0.14\textwidth}
         \footnotesize{
           $a=\{0..255\}$\\
           $\overline{t}=2,0,0$\\
           $\varphi=1$
         }
       \end{minipage}
     };
     \node[draw=blue!50!black,thick,rectangle,outer sep=5pt]
     (l22) at (4.6,-5.7) {
       \begin{minipage}{0.14\textwidth}
         \footnotesize{
           $a=\{4..255\}$\\
           $\overline{t}=2,1,0$\\
           $\varphi=1$
         }
       \end{minipage}
     };
     \node[draw=blue!50!black,thick,rectangle,outer sep=5pt]
     (l23) at (4.6,-7.1) {
       \begin{minipage}{0.14\textwidth}
         \footnotesize{
           $a=\{0..4\}$\\
           $\overline{t}=2,0,1$\\
           $\varphi=1$
         }
       \end{minipage}
     };
     \node (dots6) at (6.5,-5.7) {\textbf{$\ldots$}};
     \node (dots7) at (6.5,-7.1) {\textbf{$\ldots$}};
     \path[draw=blue!50!black] (-1.9,-7.9) rectangle (7.8,-4.9);
     \path[draw,thick,->] (top2) -- (l21);
     \path[draw,thick,->] (l21) -- (l22);
     \path[draw,thick,->] (l21) -- (l23);
     \path[draw,thick,->] (l22) -- (dots6);
     \path[draw,thick,->] (l23) -- (dots7);
     \node (A) at (8.74,-5.2) {\textcolor{blue!50!black}{set-reduced}};
     \node (A) at (8.6,-5.6) {\textcolor{blue!50!black}{transition}};
     \node (A) at (8.78,-6.0) {\textcolor{blue!50!black}{system
         $\mathcal{A}_{set}$}};
  \end{tikzpicture}
  \caption{Example verification task: for a program $P$ and a specification
    $\varphi$, the verification traverses the transition system
    $\mathcal{A}^P_{\varphi}$.}
  \label{fig:example}
\end{figure}

\begin{example} \label{exa:part}
  Consider the verification task depicted in Figure~\ref{fig:example}. The
  identification of program states can be divided into two parts: one for
  control information (marked with lighter blue in the figure) and the other for
  data (marked with darker red). Note also that the control part contains the
  program counters for individual threads of the main program and the states
  $\varphi$ of the specification automaton $\mathcal{A}_{\varphi}$. Similarly,
  it is possible to distinguish the two sources of nondeterminism in parallel
  programs: the control-flow nondeterminism (thread interleaving) is marked as
  $\zeta$-transitions and the data-flow nondeterminism (variable evaluation) as
  $\delta$-transitions.\hfill$\triangle$
\end{example}

Note that the state space of this transition system is exponential both in the
number of parallel threads and in the number of input variables. This paper
attempts to partly remedy the second state space explosion caused by the data
flow by introducing a \emph{set-based reduction}. Intuitively, the reduction
unifies those states that (1) have the same control part and at the same time
(2) the possible evaluation of their data parts form the same sets. Formally, we
can define the reduced state space inductively, starting from the initial states
$S_0=\{s^1,s^2,\ldots,s^n\}$, where $s^i=(s^i_{c},s^i_{d})$ and $\forall
i,j:s^i_c=s^j_c=s_c$. Then the one initial multi-states of the set-reduced
transition system $\mathcal{T}_{set}=(\mathfrak{S}, \rightarrow, \nu',
\mathfrak{s}_0)$ is $\mathfrak{s}_0=(s_c,\{s^i_d|i\leq n\})$. For a state
$\mathfrak{s}=(s_c,\{x_1,\ldots,x_n\})$ let $\mathsf{S}=\{s|\exists
i:(s_c,x_i)\rightarrow s\}$ be the set of successors in the unreduced state
space and $\nu':\mathfrak{S}\rightarrow 2^{\Sigma}$ have
$\nu'(\mathfrak{s})=\bigcup_{i=1}^n\nu(s_c,x_i)$. Then the successors of
$\mathfrak{s}$ in $\mathcal{T}_{set}$ form a set $\{(s,X)|x\in
X\Leftrightarrow(s,x)\in\mathsf{S}\}$.

The reduced transition system can be combined with a B{\"u}chi automaton in a
similar fashion as the unreduced state space. The resulting automaton
$\mathcal{A}_{set}=(\mathfrak{S},\rightarrow,\nu',s_0,F')$, where
$F'\subseteq\mathfrak{S}$ has the property that the set of accepting
multi-states respects the accepting states of the unreduced automaton. Formally,
let $F_{\exists},F_{\forall} \subseteq \mathfrak{S}$ such that $(s_c,X)\in
F_{\exists}\Leftrightarrow \exists x\in X:(s_c,x)\in F$ and $(s_c,X)\in
F_{\forall}\Leftrightarrow \forall x\in X:(s_c,x)\in F$. Then it holds that
$F_{\exists}=F_{\forall}$ and thus either can be used to define $F'$. The reason
for this property of the proposed reduction is that the state of the B{\"u}chi
automaton used in the product is contained in the control part of both states
and multi-states, and hence remains unreduced. Detailed reasoning would be
rather technical and the reader need only realise that while $\nu$ might
evaluate some atomic propositions differently on states within a single
multi-state, the atomic propositions used in $\varphi$ must be evaluated
consistently within a multi-state. Otherwise the respective multi-states would
be split when generating successors.

\begin{example} \label{exa:space}
  There are two programs that nicely exemplify some of the properties of
  set-based reduction, which we will use in further discussion, especially
  regarding the efficiency. Consider the following program with a loop:\\
  $\texttt{cin >> a; while ( a > 10 ) a--;}$\\
  When only the data part of multi-states is considered, the reduced transition
  system unfolds to\\
  \begin{tikzpicture}
    \node (A) at (1,0.2) {};
    \node[draw,inner sep=1pt,outer sep=3pt] (first) at (0,0) {$a=\{0..255\}$};
    \node[draw,inner sep=1pt,outer sep=3pt] (second) at (2.7,0) {$a=\{11..255\}$};
    \node[draw,inner sep=1pt,outer sep=3pt] (third) at (5.6,0) {$a=\{10..254\}$};
    \node (forth) at (7.4,0) {$\ldots$};
    \path[draw,->] (first) -- (second);
    \path[draw,->] (second) -- (third);
    \path[draw,->] (third) -- (forth);
  \end{tikzpicture}\\
  and while the state space is finite, there are as many multi-states in the
  reduction system as there were states in the original system. Furthermore,
  many states are represented multiple times: given that the first and the third
  states in the above system have the same program counter, the values of $a$
  between 10 and 254 are represented twice in these two multi-states alone.

  On the other hand, for a specification
  \begin{tikzpicture}
    \clip (0.3,-0.2) rectangle (2.5,0.2);
    \node[vertex] (empty) at (0.3,0) {};
    \node[vertex3] (fin) at (1,0) {};
    \draw (1,0) circle(2pt);
    \path[draw,->] (empty) -- (fin);
    \path[draw,->] (fin) to [in=330,out=30,min distance=10mm,loop] node
         [midway,right] {$x=1$} (fin);
  \end{tikzpicture}
  and a program\\
  $\texttt{x = 1; cin >> y; while ( true ) y++;}$\\
  the reduced transition system contains three multi-states and three
  transitions:
  \begin{tikzpicture}
     \node[draw,rectangle,outer sep=2pt]
     (first) at (0,0) {
       \begin{minipage}{0.07\textwidth}
         \footnotesize{
           $x=0$\\
           $y=0$
         }
       \end{minipage}
     };
     \node[draw,rectangle,outer sep=2pt]
     (second) at (2,0) {
       \begin{minipage}{0.07\textwidth}
         \footnotesize{
           $x=1$\\
           $y=0$
         }
       \end{minipage}
     };
     \node[draw,rectangle,outer sep=2pt]
     (third) at (4.5,0) {
       \begin{minipage}{0.14\textwidth}
         \footnotesize{
           $x=1$\\
           $y=\{0..255\}$\vspace*{-1em}
         }
       \end{minipage}
     };
     \path[draw,->] (first) -- (second);
     \path[draw,->] (second) -- (third);
    \path[draw,->] (third) to [in=350,out=10,loop] (third);
  \end{tikzpicture}\\
  whereas the original transition system contains 256 path that each enclose
  into a cycle only after 256 unfoldings of the while loop.\hfill$\triangle$
\end{example}

The set-based reduction preserves the properties of the original transition
system with respect to LTL model checking as the following theorem shows. Thus a
standard model checking procedure as described in Section~\ref{sec:pre} can be
used to verify correctness of a parallel program with respect to an LTL
property.

\begin{theorem}[Correctness]
  The product $\mathcal{A}=(S,\rightarrow,\nu,S_0,F)$ of a program transition
  system $\mathcal{T}$ and a B{\"u}chi automaton $A$ contains an accepting cycle
  iff there exists one in the reduced $\mathcal{A}_{set}=
  (\mathfrak{S},\rightarrow,\nu',\mathfrak{s}_0, F')$.
\end{theorem}

\begin{proof}
  $\Rightarrow$ Let $\leadsto$ be the reflective and transitive closure of
  $\rightarrow$. Also for any $t=(t_c,t_d)\in S$ and $\mathfrak{s}=(s_c,X)\in
  \mathfrak{S}$, let $t\sqsubseteq \mathfrak{s}$ iff $t_c=s_c\wedge t_d\in
  X\wedge \nu(t)\subseteq\nu'(\mathfrak{s})\wedge$ $t\in F\Leftrightarrow
  \mathfrak{s}\in F'$. One might observe that for any path $s_1\rightarrow
  s_2\rightarrow\ldots\rightarrow s_n$ in $\mathcal{A}$ there is a path
  $\mathfrak{s}_1\rightarrow \mathfrak{s}_2\ldots\rightarrow \mathfrak{s}_n$ in
  $\mathcal{A}_{set}$ such that for all states along the path it holds that
  $s_i\sqsubseteq\mathfrak{s}_i$. Assume $t_0\leadsto t_{1}\rightarrow\ldots
  \rightarrow t_{n}=t_{1}$ in $\mathcal{A}$ and hence also $\mathfrak{t}_0
  \leadsto \mathfrak{t}_{1}\rightarrow\ldots\rightarrow \mathfrak{t}_{n}$ in
  $\mathcal{A}_{set}$. But it the reduced state space it might happen that
  $\mathfrak{t}_n\neq\mathfrak{t}_1$ but instead $\mathfrak{t}_n=
  \mathfrak{t}'_1$. Unrolling the cycle in $\mathcal{A}$ further we get
  $\mathfrak{t}_1\leadsto\mathfrak{t}'_1\leadsto\mathfrak{t}''_1 \leadsto\ldots$
  in $\mathcal{A}_{set}$. Yet if $\mathfrak{t}=(t_c,X)$ then $t_{c1}=t'_{c1}=
  \ldots$ and also $X_1\supseteq X'_1\supseteq\ldots$. To understand why the
  second part of the previous statement holds one needs only to remember that
  given $t_1=t_n$ the combined effect of the program between $t_1$ and $t_n$ on
  the data part has to be an identity. $X_1, X'_1,\ldots$ may not be equal only
  because the program conditions along the path further limit what values of
  input variables might have led to this state of computation. It immediately
  follows that the sequence $X_1,X'_1,\ldots$ has a fix point
  $\mathfrak{t}^m_1$, which is the first multi-state of a cycle in
  $\mathcal{A}_{set}$. Finally, since the relevant path in $\mathcal{P}$ may be
  arbitrarily unrolled along the cycle $t_1\leadsto t_1$, it still holds that
  for $t_0\leadsto t_1(\leadsto t_1)^m\leadsto t_1$ and $\mathfrak{t}_0 \leadsto
  \mathfrak{t}^m_1\leadsto\mathfrak{t}^m_1$ $t\sqsubseteq\mathfrak{t}$ along the
  paths and thus the cycle $\mathfrak{t}^m_1\leadsto\mathfrak{t}^m_1$ is
  accepting in $\mathcal{A}_{set}$.

  $\Leftarrow$ Assume $\mathfrak{t}_0\leadsto\mathfrak{t}_1\rightarrow\ldots
  \rightarrow\mathfrak{t}_n=\mathfrak{t}_1$ in $\mathcal{A}_{set}$. Then as
  above there must be a path $t_0\leadsto t_1\rightarrow\ldots\rightarrow t_n$
  in $\mathcal{A}$ such that $t\sqsubseteq\mathfrak{t}$ all along the path but
  again $t_1$ may not be equal to $t_n=t'_1$. Let $t_1=(t_{c1},t_{d1})$ and let
  $\kappa$ be the operation applied on $t_{d1}$, i.e. $t'_{d1}= \kappa(t_{d1})$,
  then $t'_{c1}=t''_{c1}=\ldots$ and we will show that there exists $m$ such
  that $\kappa^m(t_{d1})=t_{d1}$. It follows from the fact that the underlying
  structure of the data part is a commutative ring $\mathbb{Z}/k\mathbb{Z}$ of
  integers modulo $k$, where $k$ is the product of the domains of input
  variables. Computer programs use modular arithmetic and it is a property of
  such arithmetic that for any operation $\kappa$ there is an $m$ such $m$
  iterations of $\kappa$ is an identity. The rest of this implication is similar
  to the previous one.\qed
\end{proof}

As apparent from Example~\ref{exa:space}, reasoning about the efficient of the
proposed set-based reduction -- the ratio between the size of the original
system and the size of the reduced system -- is rather complicated. For a
program without cycles, the reduction is exponential with respect to the number
of input variables and to the sizes of their domains. Note, however, that for
trivial cases of data-flow nondeterminism even this reduction can be
negligible. The case of programs with cycles is considerably more involved.

Let us call cycles those paths in a transition system that start and end in two
states with the same control part, $s=(s_c,s_d)$ and $s'=(s_c,s_d')$. Then the
function $f$ of the cycle, transforming $s_d$ to $s_d'$, has a fix point as was
argued in the above proof, and this fix point has to be computed (explicitly in
our case, as opposed to symbolic solution~\cite{Lin96} of the same
problem). That aspect is present in full and reduced state spaces alike, yet may
produce an exponential difference in their sizes. If the multi-state already
contains the fix point before it reaches given cycle, as in the second program
of Example~\ref{exa:space}, then the reduced system contains only as many
multi-states as is the length of that cycle. On the other hand, as the first
program of Example~\ref{exa:space} demonstrated, the reduction can even be to
the detriment of the space complexity even if we assume that the size of
multi-states is sublinear in the number of states contained within, which is
difficult to achieve, as we discuss in the conclusion.

The remainder of this section will investigate the necessary properties the
hybrid representation must possess to enable LTL model checking by
following the steps forming the model checking process and describing how each
must be altered. As described in Section~\ref{sec:pre}, LTL model checking
requires implementation of three functions: \textsf{initial state},
\textsf{successors}, and \textsf{is accepting}.

\subsection{Changes in LTS Traversal}

\paragraph{Initial Ranges}
The multi-states, as mentioned above, consist of two parts, but unless it proves
to impede the clarity we will assume only the symbolic part to be present. Under
this assumption the \textsf{initial state} function returns a set of every
combination within the initial ranges of the undefined variables. For example,
in case of two variables $a\in\{1,\ldots,5\}$, and $b\in\{4,\ldots,5\}$, the
initial multi-state represents a set $\{(a,b)|1\leq a\leq 5\wedge 4\leq b\leq
5\}$.

\paragraph{Assignments and Conditions}
Generating successors must take into consideration the branching of control flow
and must allow changing the evaluation of variables. Without the loss of
generality one can expect the \textsf{successors} to use only two methods to
interact with variables: \textsl{prune} and \textsl{apply}. \textsl{prune} takes
a Boolean expression $e$, evaluates it and removes all evaluations in the
multi-state that do not satisfy $e$. \textsl{apply} takes an assignment, a pair
(variable $v$, expression $e$) and updates the evaluation accordingly. Applying
an assignment on a multi-state entails considering every combination of stored
values, evaluating $e$ on that combination and finally updating $v$ for the
value of $e$. Conditional branching is handled by \textsl{prune} and assignments
are handled by \textsl{apply}. Which leaves only cycles.

\paragraph{Decidable Equality}
Dealing with cycles represents a major problem for execution-based
verification. They are either unwound~\cite{BDK08}, which is imprecise, or
considered naively, which leads to infinite state spaces~\cite{KPV03}. Our
insisting on having LTL specifications, however, has one very specific
consequence when dealing with cycles. Accepting cycle detection algorithms
require duplicate detection to be decidable, i.e. the representation must enable
checking equality of multi-states. Hence every multi-state is stored only once
and consequently, the state space must be finite, even with cycles in the LTS.

It might appear that differentiating every two multi-state that only differ in
their data parts produces unnecessarily too large state spaces; that
subsumption~\cite{XMS05} could be a sufficient condition for state
equality. That is not correct with respect to LTL. For a state $S$ assume that a
different state $S'$ is found such that $S'\subseteq S$. If these were matched
into a single state $S$ and there was a path from $S'$ to $S''\subseteq S$ such
that $S''\cap S'=\emptyset$, then the reduced transition system would contain a
cycle where there was none in the original system.

\subsection{Changes in Counterexample Computation}\label{sec:chaCou}

Pruning when generating successors leads to complications because the current
multi-state is implicitly divided, based on which of its evaluations satisfy
given condition. However, the information necessary for such division is only
available when the successors are generated, i.e. after the actual source
multi-state was stored. Only with hindsight can one express what evaluation of
the input variables leads to a certain state.

This unintended consequence does not affect the model checking procedure itself,
but it affects its crucial part: the counterexample generation. A counterexample
represents the path that leads to an accepting cycle -- the piece of information
that specifies the defect of the system under verification. Standard explicit
model checking generates counterexamples by traversing the LTS backwards from
the accepting cycle along the so-called \emph{parent graph}, a tree generated
during the forward traversal. To remedy this consequence of using multi-states,
it suffices (during the backwards traversal) to prune the multi-state to contain
only the correct evaluations. An example: at some point the backwards traversal
follows a transition that leads to a multi-state, which, in order to follow the
transition in forward traversal, must satisfy $a>5$. Then those evaluations for
which $a\leq 5$ are removed from this multi-state. Note also that this approach
is robust even to accommodate the reduced cycle of the second program in
Example~\ref{exa:space}. 

\section{Case Study}

In this introductory paper we aim at validating the proposed method and use
explicit sets to represent multi-states. There still are space reductions,
considerable as will be demonstrated in the experiments, resulting from the
redundancy exhibited by the repeated execution. The evaluation of defined
variables and control-flow information, e.g. the program counter, was stored in
every state, once for every evaluation of undefined variables. Now it is only
stored once, as the explicit part of a multi-state. The space reductions would
be undeniably greater have we used a symbolic representation of sets. However,
as mentioned before, LTL specification is paramount to us, equality of
multi-states must be decidable -- a property which many symbolic representations
lack.

\subsection{The DVE Language}

The DVE language was established specifically for the design of protocols for
communicating systems. There are three basic modelling structures in DVE:
\texttt{processes}, \texttt{states}, and \texttt{transitions}. At any given
point of time, every \texttt{process} is in one of its \texttt{states} and a
change in the system is caused by following a \texttt{transition} from one
\texttt{state} to another. Communication between processes is facilitated by
global variables or \texttt{channels}, that connect two \texttt{transitions} of
different \texttt{processes}: the two \texttt{transitions} are followed
concurrently. Following a \texttt{transition} is conditioned by \texttt{guard}
expressions, that the source \texttt{state} must satisfy, and entails
\texttt{effects}, an assignment modifying the variable evaluation. LTL
specification is merely another \texttt{process}, whose \texttt{transitions} are
always connected to the system \texttt{transitions}. Comprehensive treatment of
the DVE syntax and semantics can be found in~\cite{Sim06}.

The DVE language allows using variables of different types and consequently of
different sizes and, thus, the proposed modification only adds the specification
which variables are undefined and what are their ranges. States are represented
as an evaluation of variables and a map assigning a \texttt{state} to each
\texttt{process}. In multi-states, the original states are preserved as the
explicit part and the symbolic part is a representation of the undefined
variables (stored as a new, defined variable in the explicit part). Once
restored from memory, the explicit evaluation of variables forms the so-called
\emph{context}.

When following a \texttt{transitions}, the function \textsf{successors} calls
the method \textsl{prune} with a \texttt{guard} as its parameter. If the
representation results in being empty, i.e. not a single evaluation satisfies
the \texttt{guard}, then no successors are generated. Otherwise, the
\texttt{effect} is \textsl{applied} and the resulting representation is stored
in a new multi-state. The evaluation of expressions is undertaken in a standard
way, except that every combination of undefined variables is first loaded into
the context: there is no need to modify the underlying arithmetic.

\subsection{Experiments}\label{sec:exp}

The above described case study for the DVE language was
implemented~\footnote{Code available at
  {\footnotesize\url{http://anna.fi.muni.cz/~xbauch/code.html\#symbolics}}.} in
the DiVinE~\cite{BBH13} verification environment, which already supported the
DVE input language for system description. The change consisted only of the
addition of hybrid representation, and by extension of support for partially
open systems. The model checker remained unmodified, the parallel accepting
cycle detection algorithms and specialised data structures could still be used
without additional alternation of the code. Similar conditions were chosen for
the comparison between repeated execution (unmodified DiVinE) and the new hybrid
approach: the codes were compiled with optimisation option \texttt{-O2} using
GCC version 4.7.2 and ran on a dedicated Linux workstation with 64 core Intel
Xeon 7560 @ 2.27GHz and 512GB RAM.

\begin{figure}[t]
  \begin{subfigure}{0.49\textwidth}
    \includegraphics[width=\textwidth]{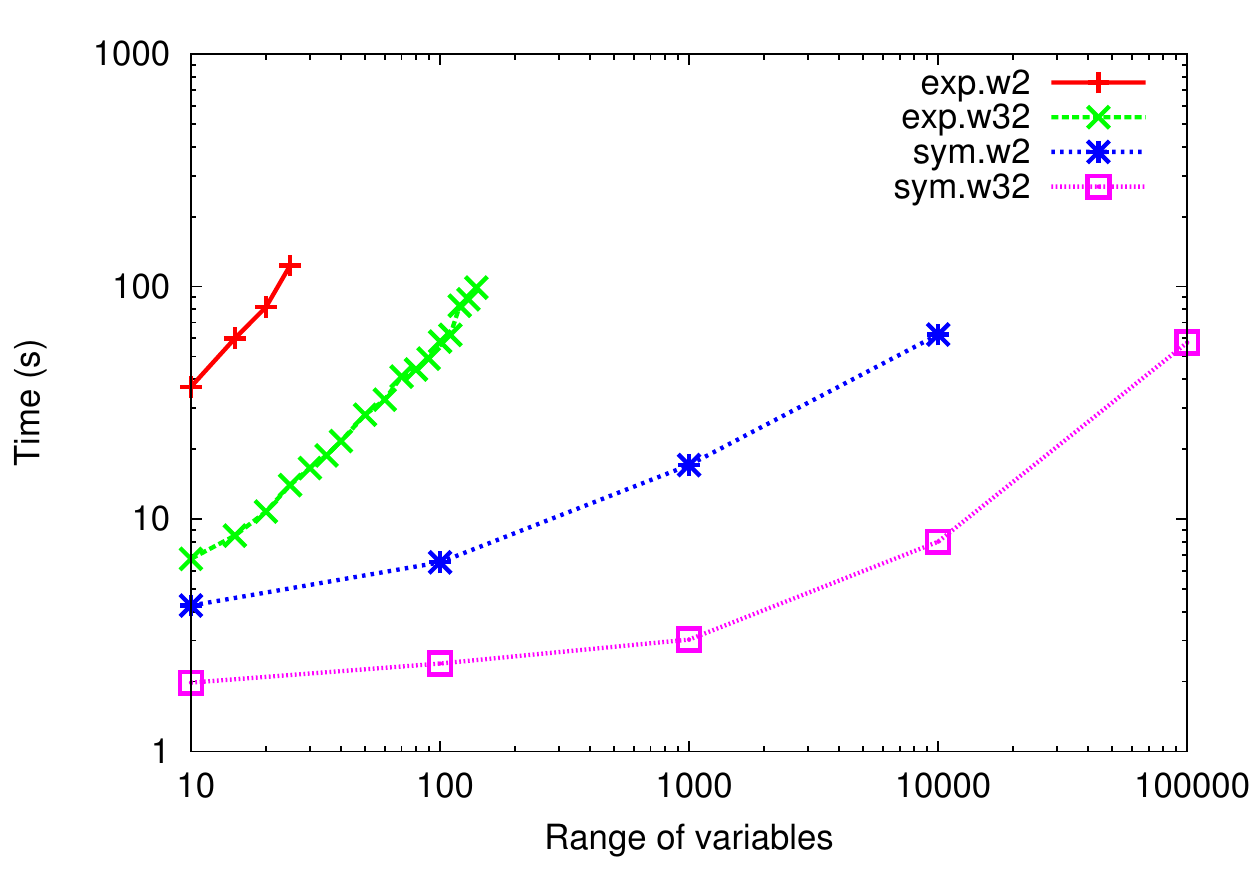}
    \caption{\small Time complexity for peterson-liveness}
    \label{fig:time}
  \end{subfigure}
  \begin{subfigure}{0.51\textwidth}\vspace*{-0.2em}
    \includegraphics[width=\textwidth]{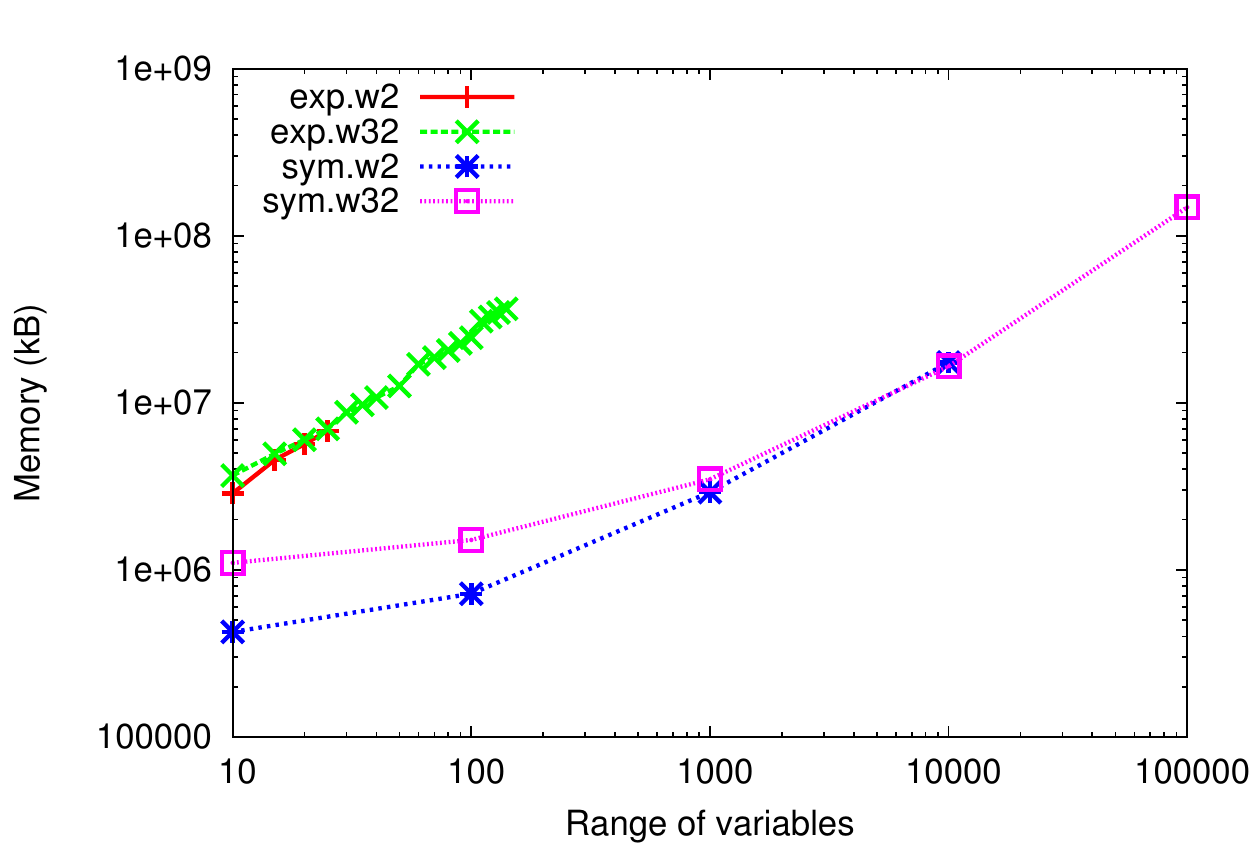}
    \caption{\small Space complexity for peterson-liveness}
    \label{fig:space}
  \end{subfigure}
\end{figure}

We have conducted a set of experiments pertaining to the Peterson's
communication protocol. For the purposes of verification, the protocol is
usually modelled in such a way that once a process accesses the critical
section, it immediately leaves the critical section, without performing any
work. The introduction of input variables allows the model to achieve closer
approximation of practical use by simulating some action in the critical
section, however artificial that action might be. Hence a global input variable
$l\in\{0\ldots \mathsf{r}\}$ was added to the model and an action
$l=(l+1)\%\mathsf{r}$. Note that the action is not biased towards set-based
reduction because it forces inclusion of all subsets $\{0\ldots\mathsf{r}\},
\{1\ldots\mathsf{r}\},\ldots$ even in the reduced state space.

The two plots above report the results of liveness verification of this modified
Peterson's protocol. Verification of this protocol is nontrivial and the best
parallel algorithm OWCTY (see~\cite{BBH13} for more details of this and other
parallel algorithms used in DiVinE) requires several iteration before it can
answer the verification query. The experiments were executed using the fully
explicit approach of~\cite{BBB12c}, denoted as \textsf{exp}, and the hybrid
approach proposed in this paper, denoted as \textsf{sym}; using 2 and 32
parallel threads (\textsf{w2} and \textsf{w32}). The plots clearly show that the
fully explicit approach cannot scale with the range of input variables $r$ (the
$x$-axis) and even when 32 parallel threads were used, verification of a single
variable of range $0..140$ required almost 100 seconds. Our hybrid approach scaled
markedly better, easily achieving the range up to 10000 with the same spacial
complexity that \textsf{exp} needed for two orders of magnitude smaller range.

\section{Conclusion}

This paper represents an initial step towards complete and precise verification
of parallel software against temporal specification. We investigate the
potential of the combination of explicit and symbolic approaches, handling the
control-flow explicitly and the data-flow symbolically, as the mean of taking
this step. The potential is demonstrated in the preliminary results, on the
experiments conducted with a communication protocol and a trivial explicit set
representation, where the scalability of combining explicit and symbolic
approaches surpasses the purely explicit approach. Even with the most basic
symbolic representation for data, we have multiplied the allowed range of input
variables. The data domain is still bounded, but that is a reasonable price to
pay for temporal specification: one only needs to expand the boundary from
$0$--$n$ to $0$--$2^n$.

Moving from a linear bound to a logarithmic bound on data is the long-term goal
of our research. Purely symbolic representations (BDDs and similar) might allow
such a move, but these are limited as to what operations on the data the
representations support. More immediate possibilities lie in relaxing some of
the imposed limitations, e.g. supporting only Presburger arithmetic would still
allow precise verification on logarithmically bounded variables, while retaining
the ability to verify against temporal specification. The first-order theory of
bit-vectors appears most promising; there the greatest challenge would be the
methodology of comparing two multi-states and the ranges of input variables much
manageable problem.

\bibliographystyle{plain}
\bibliography{main}

\end{document}